# Growth of Smaller Grain Attached on Larger One: Algorithm to Overcome Unphysical Overlap between Grains

Acep Purqon[1,a)] and Sparisoma Viridi[2,b)]

[1]*Earth Physics and Complex System Research Division,
Institut Teknologi Bandung, Jalan Ganesha 10, Bandung 40132, Indonesia*
[2]*Nuclear Physics and Biophysics Research Division,
Institut Teknologi Bandung, Jalan Ganesha 10, Bandung 40132, Indonesia*

[a)]acep@fi.itb.ac.id
[b)]Corresponding author: dudung@fi.itb.ac.id

**Abstract.** As a smaller grain, which is attached on larger one, is growing, it pushes also the larger one and other grains in its surrounding. In a simulation of similar system, repulsive force such as contact force based on linear spring-dashpot model can not accommodate this situation when cell growing rate is faster than simulation time step, since it produces sudden large overlap between grains that makes unphysical result. An algorithm that preserves system linear momentum by introducing additional velocity induced by cell growth is presented in this work. It should be performed in an implicit step. The algorithm has successfully eliminated unphysical overlap.

## INTRODUCTION

Simulation of grains interaction can be used to observe cell swap in caging effect [1], biological cell sorting [2], and tissue growth [3], where the latest is similar to colony growth as observed in [4]. One of the problems in simulating this kind of interaction during growing process of a grain is that the produced overlap between grains is increasing faster than the changing position of the interacting grains, which can introduce computation failure. It is the reason why only small grains overlap considered [3]. In this work an implicit step is proposed, in where grains rearrangement is taking place but without changing simulation time step. In the algorithm linear momentum is conserved. For now only example of two grains interaction is discussed.

## THEORY

Supposed that there are two grains $i$ and $j$, each with radius $R_i$ and $R_j$ and located at position $\vec{r}_i$ and $\vec{r}_j$, respectively. Overlap between these two grains can be defined as [5]

$$\xi_{ij} = \max\left(0, R_i + R_j - |\vec{r}_i - \vec{r}_j|\right), \qquad (1)$$

where

$$\max(a,b) = \begin{cases} a, & a > b, \\ b, & a < b. \end{cases} \qquad (2)$$

Assumed that grain $j$ has already existed and grain $i$ is grown on grain $j$ at position

$$\vec{r}_i = \vec{r}_j + R_j \hat{r}_{ij} \qquad (3)$$

with

$$\hat{r}_{ij} = \frac{\vec{r}_i - \vec{r}_j}{|\vec{r}_i - \vec{r}_j|}. \qquad (4)$$

Radius of grain $i$ is defined as

$$R_i(t) = \begin{cases} 0, & t < t_i^+, \\ \dot{R}_i^+ & t_i^+ \leq t \leq t_i^+ + T_{ji}^+, \\ R_{max}, & t > t_i^+ + T_i^+, \end{cases} \qquad (5)$$

where $t_j^+$ is birth time and $T_i$ is mature time of grain $i$, respectively. $\dot{R}_i^+$ stands for radius growth rate for grain $i$ and $R_{max}$ maximum radius of every grain. Equation (3) holds at $t = t_i^+$ and according to Equation (5) overlap between grains $i$ and $j$ is produced using Equation (1) as illustrated in Fig. 1(b) below.

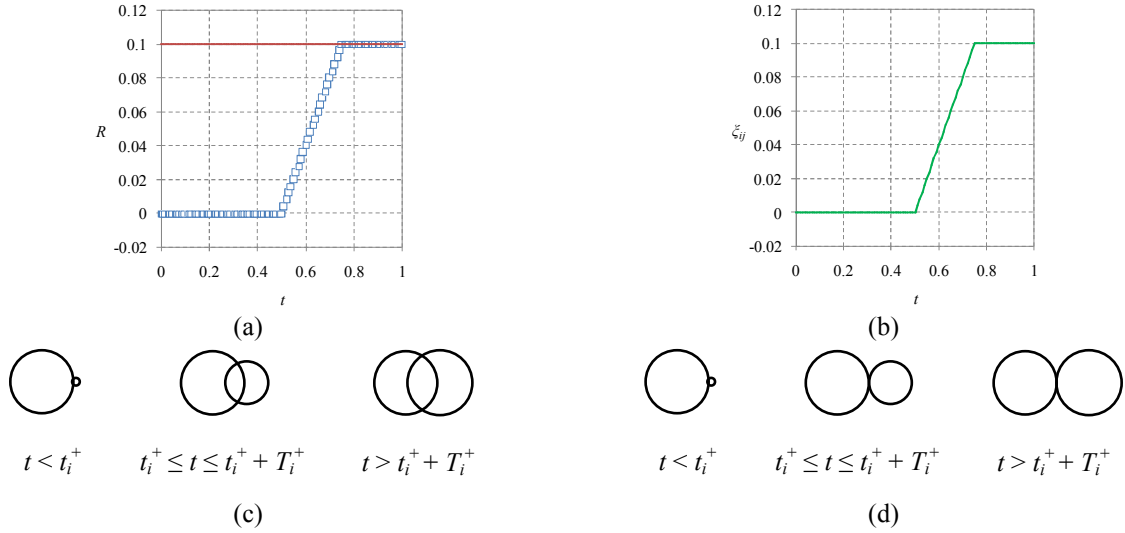

**FIGURE 1.** (a) Grains radius $R_i$ (empty box) and $R_j$ (solid line) as function of time for $\dot{R}_i^+ = 0.4$, $t_i^+ = 0.5$, $T_i^+ = 0.25$, $R_{max} = 0.1$; (b) Overlap between grains $i$ and $j$ due to radius growth of grain $i$ as function of time; (c) Unphysical grain overlap; (d) Expected and physical grain overlap.

Figure 1 tells us that this overlap $\xi_{ij}$ is not physically correct since there is no external force to maintain the overlap. Then, there should be a mechanism to move the grains apart, while conserving linear momentum of the two grains system. In other word, Fig. 1(d) gives the desired condition similar as shown in an observation video [4], while in a reported simulation, the condition in Fig. 1(c) is avoided by limiting the system only to small overlap [3].

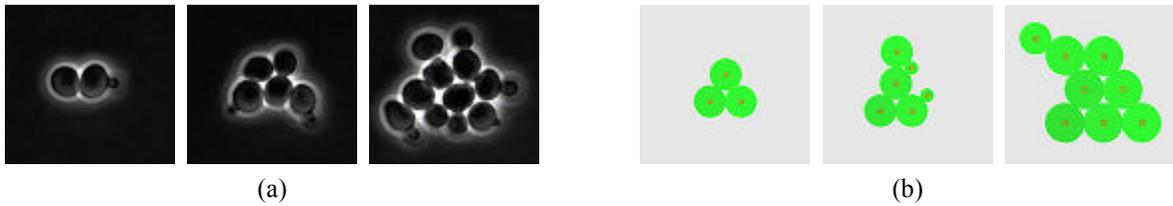

**FIGURE 2.** Growth of: (a) nearly spherical grains of yeast as observed [4] and (b) 2-d sperical grains in simulation for small overlap condition [3].

It is also assumed that every grain has the same mass density σ

$$\sigma = \frac{m}{\pi R_{max}^2},\qquad(6)$$

that defines mass of the growing grain

$$m(t) = \pi \sigma R_i^2(t).\qquad(7)$$

System center of mass will be

$$\vec{r}_{COM} = \frac{m_i(t)\vec{r}_i + \sum_{j=1}^{N} m\vec{r}_j \delta_{ij}}{(N-1)m + m_i(t)}.\qquad(8)$$

## Two grains system

For system consists only of two grains $i$ and $j$, their velocity can be defined through the process of minimizing overlap $\xi_{ij}$. Let assume that at each simulation time step $\Delta t$ value of the overlap should be suddenly set to zero through

$$\vec{v}_i = \left[\frac{m}{m + \pi \sigma R_i^2(t)}\right]\frac{\xi_{ij}}{\Delta t}\hat{r}_{ij}.\qquad(9)$$

and

$$\vec{v}_j = \left[\frac{\pi \sigma R_i^2(t)}{m + \pi \sigma R_i^2(t)}\right]\frac{\xi_{ij}}{\Delta t}\hat{r}_{ji}.\qquad(10)$$

The algorithm to change grains position is quite simple, that every grain should be moved to opposite direction using Euler method

$$\vec{r}_i(t) = \vec{r}_i(t) + \vec{v}_i \Delta t,\qquad(11)$$

but it should be conducted in implicit step instead in the real simulation step, which means that before overlap eliminated completely, explicit step can not be performed. Equations (9) and (10) will guarantee momentum conservation

$$m\vec{v}_j + \pi \sigma R_i^2(t)\vec{v}_i = 0,\qquad(12)$$

since $\hat{r}_{ji} = -\hat{r}_{ij}$.

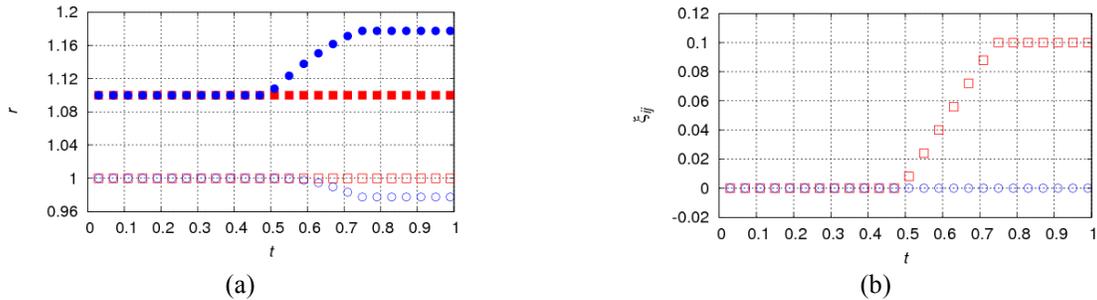

**FIGURE 3.** (a) Position of grains $i$ and $j$ before (square) and after (circle) using implicit step; (b) Overlap between grains $i$ and $j$ before (square) and after (circle) using implicit step.

## RESULTS AND DISCUSSION

Position of grains $i$ and $j$ and also their overlap $\xi_{ij}$ is given in following Fig. 3. It is shown in Fig. 3(b) that the implicit step successfully eliminates unphysical overlap (empty circle marker). Parameters values used in the simulation are the same as in caption of Fig. 1, if not explicitly differently stated. At $t = 0.5$ after using the implicit step grain $i$ starts to go to positive direction while grain $j$ goes to opposite direction. This process occurs only during mature time $T_i^+ = 0.25$ and then they stop moving. Then after $t = 0.75$ both grains remain at same place.

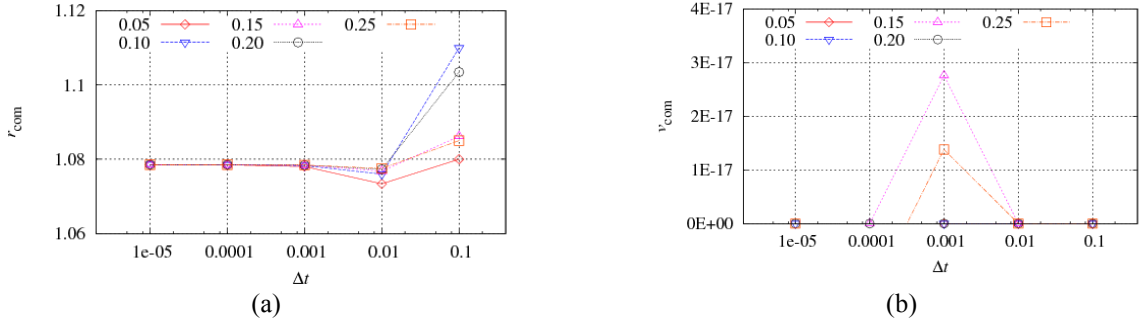

**FIGURE 4.** (a) Position of center of mass as function of $\Delta t$ for different value of $T_i^+ = 0.05 \cdots 0.25$; (b) Velocity of center of mass as function of $\Delta t$ for different value of $T_i^+ = 0.05 \cdots 0.25$. Initial position of grain $j$ is at $r = 1$.

Later exploration of the algorithm shows that value of $\Delta t = 10^{-5}$ already gives consistent results both for position and velocity of center of mass, which is independent of mature time $T_i^+$ as shown in Fig. 4. Unfortunately, there is still unphysical shift of the two grain system about 0.08 from initial position of grain $j$ at $r = 1$, which is about 8 % error. It seems the accuracy limit of this algorithm since smaller value of $\Delta t$ can not produce better position of center of mass. Though, the velocity center of mass already produce desired result for $\Delta t = 10^{-5}$.

## SUMMARY

An algorithm, which is performed in implicit step to overcome unphysical overlap between two grains during grain growing, has been discussed. It has successfully eliminated the overlap while maintaining momentum conservation, but unfortunately there is still an error since unphysical shift of center of mass position occurs.

## ACKNOWLEDGEMENTS

Authors would like to thank to ITB Research Grant in year 2014 (Contract Number 914/AL-J/DIPA/PN/SPK/2014 and XYZ/AL-J/DIPA/PN/SPK/2014) supporting this work and FMIPA ITB for supporting the presentation.

## REFERENCES

1. E.-M. Schötz, M. Lanio, J. A. Talbot, and M. L. Manning, J. R. Soc. Interface **10**, 20130726 (2013).
2. F. Graner and J. A. Glazier, Phys. Rev. Lett. **69**, 2013-2016 (1992).
3. D. Aprianti, L. Haris, F. Haryanto, S. N. Khotimah, and S. Viridi, "Two-dimension Tissue Growth Model Based on Circular Granular Cells for Cells with Small Overlap," in International Symposium on Computational Science-2014, Bandung, Indonesia, 20 May 2014, edited by M. A. Martoprawiro et al. (Institut Teknologi Bandung and Kanazawa University), *in press*; arXiv:1403.6784v1 [physics.bio-ph] Wed, 26 Mar 2014.
4. BucherLab, "Timelapse Movie of Fluorescent Yeast", YouTube, 14 Nov 2009, URL http://www.youtube.com /watch?v=FcV1ydls9hg [20140820].
5. J. Schäfer, S. Dippel, and D.E. Wolf, J. Phys. I France **6**, 5-20 (1996).

# APPENDIX

```cpp
/*
    2govelim.cpp
    Two grains overlap elimination
    Sparisoma Viridi | dudung@gmail.com
    
    Compile: g++ govelim.cpp -o govelim
    Execute: ./govelim
    
    20141101 Start writing this code since spreadsheet version
             failed
*/

#include <iostream>
#include <fstream>
#include <cmath>

using namespace std;

int main(int argc, char *argv[]) {
    // Verbose program name
    const char *pname = "2govelim";
    cout << pname << ":" << endl;
    
    // Define some constants
    const char *tab = "\t";
    const double Rmin = 0;
    const double Rmax = 0.1;
    const double dt = 1E-1;
    const double tp = 0.5;
    const double Tp = 0.20;
    const double dRp = Rmax / Tp;
    const double ti = 0;
    const double tf = 1;
    const double sigma = 1;
    const double eps_xi = 1E-10;
    const double alpha = 1.0;
    const bool IMPLICIT_STEP = true;
    
    // Define array for particle
    const unsigned int N = 2;
    double m[N] = {M_PI * Rmax*Rmax * sigma, 0};
    double R[N] = {Rmax, Rmin};
    double r[N] = {1, 1 + R[0]};
    double v[N] = {0, 0};
    
    // Display output header
    cout << "# t\tR_0\tR_1\tr_0\tr_1\t\\xi_{01}\tr_{com}";
    cout << "\tN_{step}\tv_0\tv_1\tv_{com}";
    cout << endl;
    
    // Perform iteration
    double t = ti;
    while(t < tf + dt) {
        // Grow radius of grain 1 and its mass
        if(tp <= t && t <= tp + Tp) {
            R[1] += dRp * dt;
        }
        m[1] = M_PI * R[1]*R[1] * sigma;
        double M = m[0] + m[1];
        
        // Calculate overlap
        double buf = R[0] + R[1] - fabs(r[0] - r[1]);
        double xi = 0 > buf ? 0 : buf;
        
        // Calculate rcom
        double rcom = (m[0] * r[0] + m[1] * r[1]) / M;
```

```cpp
            double rcom0 = rcom;

            // Perform implicit step
            unsigned int istep = 0;
            while(xi > eps_xi && IMPLICIT_STEP) {
                // Calculate velocity
                v[0] = -m[1] / M * (alpha * xi / dt);
                v[1] = m[0] / M * (alpha * xi / dt);

                // Change position
                r[0] += v[0] * dt;
                r[1] += v[1] * dt;

                // Calculate overlap
                buf = R[0] + R[1] - fabs(r[0] - r[1]);
                xi = 0 > buf ? 0 : buf;

                istep++;
            }

            // Calculate rcom again
            rcom = (m[0] * r[0] + m[1] * r[1]) / M;
            double vcom = (m[0] * v[0] + m[1] * v[1]) / M;

            // Display value of parameters
            cout << t << tab;
            for(unsigned int i = 0; i < N; i++) {
                cout << R[i] << tab;
            }
            for(unsigned int i = 0; i < N; i++) {
                cout << r[i] << tab;
            }
            cout << xi << tab;
            cout << rcom << tab;
            cout << istep << tab;
            for(unsigned int i = 0; i < N; i++) {
                cout << v[i] << tab;
            }
            cout << vcom << endl;

            // Increase simulation time by a time step
            t += dt;
        }

        return 0;
}
```